\documentclass[twocolumn,showpacs,amsmath,amssymb,prb,superscriptaddress]{revtex4-1}
\usepackage{graphicx}
\usepackage{color}

\begin{document}


\title{Effective g factor of {2D holes} in strained {Ge} quantum wells}

\author{I.~L.~Drichko}
\affiliation{Ioffe Institute, 26 Politekhnicheskaya, 194021 St.~Petersburg, Russia}

\author{A.~A.~Dmitriev}
\affiliation{Ioffe Institute, 26 Politekhnicheskaya, 194021 St.~Petersburg, Russia}
\affiliation{Department of~Nanophotonics and Metamaterials, ITMO University, 49 Kronverksky Pr., 197101 St.~Petersburg, Russia}

\author{V.~A.~Malysh}
\affiliation{Ioffe Institute, 26 Politekhnicheskaya, 194021 St.~Petersburg, Russia}

\author{I.~Yu.~Smirnov}
\affiliation{Ioffe Institute, 26 Politekhnicheskaya, 194021 St.~Petersburg, Russia}

\author{H.~von~K{\"a}nel}
\affiliation{Laboratory for Solid State Physics, ETH Zurich, Otto-Stern-Weg 1, CH-8093 Zurich, Switzerland}

\author{M.~Kummer}
\affiliation{Laboratory for Solid State Physics, ETH Zurich, Otto-Stern-Weg 1, CH-8093 Zurich, Switzerland}

\author{D.~Chrastina}
\affiliation{INFM and L-NESS Dipartimento di~Fisica, Politecnico di~Milano, Polo Regionale di~Como, Via Anzani 52, I-22100 Como, Italy}

\author{G.~Isella}
\affiliation{INFM and L-NESS Dipartimento di~Fisica, Politecnico di~Milano, Polo Regionale di~Como, Via Anzani 52, I-22100 Como, Italy}

\date{\today}

\begin{abstract}
The effective g-factor of 2D holes in modulation doped \mbox{p-SiGe/Ge/SiGe}
structures was studied. The AC conductivity of samples with hole densities from
$3.9 \times 10^{11}$~to $6.2 \times 10^{11}~\text{cm}^{-2}$ was measured in
perpendicular magnetic fields up to $8~\text{T}$ using a contactless acoustic
method. From the analysis of the temperature dependence of conductivity
oscillations, the $\mathrm{g}_{\perp}$-factor of each sample was determined.
The $\mathrm{g}_{\perp}$-factor was found to be decreasing approximately
linearly with hole density.  This effect is attributed to non-parabolicity of
the valence band.
\end{abstract}


\maketitle

\section{Introduction}
Over the last few years, modulation doped \mbox{SiGe/Ge/SiGe} structures with a
2D hole gas have constantly been a subject of interest, due to their high hole
mobilities and epitaxial compatibility with Si. Recently, homogenous samples
with hole mobility high enough to observe the integer and even the fractional
quantum Hall effect have become available.\cite{ZudovFQHE,MironovFQHStates}

In such structures the lattice constant mismatch between SiGe and Ge leads to
compressive strain in the Ge quantum well.
 This strain lifts the degeneracy of the heavy hole and
light hole subbands in the $\Gamma$-point.
In the relevant case of compressive strain, the heavy hole subband lies higher
in energy, which gives the opportunity to investigate the properties of only
heavy holes.  The subbands, interact with each other, however, and this
interaction gives rise to valence band non-parabolicity. It can manifest itself
through hole density dependencies of effective mass and g-factor. The former
has been studied extensively.\cite{MironovIrisawa,HansRossner2007}

Because of the large g-factor of bulk Ge ($\mathrm{g} = 20.4$), the
$\mathrm{g}_{\perp}$-factor of 2D holes in Ge quantum wells is also expected to
be large and to depend on the hole density due to valence band
non-parabolicity.  However, until very recently, only a few values of the
$\mathrm{g}_{\perp}$-factor in \mbox{SiGe/Ge/SiGe} structures have been present
in literature.\cite{ZudovFQHE,Arapov,ContributionsGFactor,Nenashev} The first
investigation of the $\mathrm{g}_{\perp}$-factor dependence on hole density was
published in 2017.\cite{SmallSheetDensityGFactor} Its authors studied the
the $\mathrm{g}_{\perp}$-factor of 2D holes in a heterostructure field-effect
transistor with low hole densities from $1.4 \times 10^{10}$~to $1.3 \times
10^{11}~\text{cm}^{-2}$. The aim of the present work is to continue the
investigation of the $\mathrm{g}_{\perp}$-factor in the region of higher hole
densities.

\section{Experimental results and discussion}
\subsection{Samples}
We have studied modulation-doped
single
\mbox{p-Si$_x$Ge$_{1-x}$/Ge/Si$_x$Ge$_{1-x}$}
 quantum wells with hole densities from $3.9 \times 10^{11}$~to $6.2 \times
10^{11}~\text{cm}^{-2}$. They had all been grown on Si(001) substrates using
low-energy plasma-enhanced chemical vapor deposition
(LEPECVD).\cite{LEPECVDforSiGe} Symmetrically and asymmetrically doped samples
have both been studied. A summary of the samples is presented in
Table~\ref{tab:Samples}.

\begin{table}[bt]
  \caption{\label{tab:Samples} Summary of the samples.}
  \centering
  \begin{tabular}{| c c c c c |}
    \hline
    Sample & Hole density & QW width & $x$ & Doping layers \\
     & ($\text{cm}^{-2}$) & ($\text{nm}$) & & \\\hline
    7989-2 & $3.9 \times 10^{11}$ & 13 & 0.3 & at each side \\
    6745-3 & $4.5 \times 10^{11}$ & 15 & 0.3 & single-sided \\
    8005   & $4.5 \times 10^{11}$ & 13 & 0.3 & at each side \\
    6745-2 & $4.9 \times 10^{11}$ & 15 & 0.3 & single-sided \\
    6777-2 & $5.2 \times 10^{11}$ & 15 & 0.3 & single-sided \\
    K6016  & $6.0 \times 10^{11}$ & 20 & 0.4 & single-sided \\
    7974-2 & $6.2 \times 10^{11}$ & 13 & 0.3 & at each side \\
    \hline
  \end{tabular}
\end{table}

\subsection{Method}
We used the contactless acoustic
method\cite{WixforthPioneer,K6016AcoustoElectronicEffects} to determine the
complex AC conductance of the 2D hole gas, $\sigma = \sigma_1 - i\sigma_2$. In
the present work, however, we were generally focused on the real part,
$\sigma_1$, of the complex conductance.

\begin{figure}[hbt]
  \centering
  \includegraphics[width=8.5cm, height=4cm, keepaspectratio]{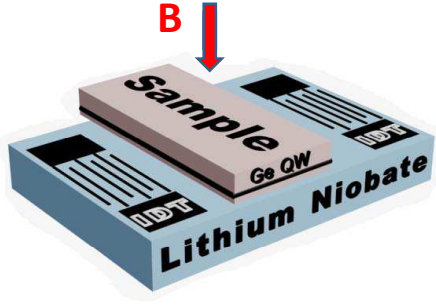}
  \caption{\label{fig:Schematic} A schematic of the experimental setup.}
\end{figure}

A schematic of the experiment setup is shown in Fig.~\ref{fig:Schematic}.  The
samples are mounted onto a \mbox{LiNbO$_3$} piezoelectric crystal with a pair
of interdigital transducers (IDT), which induce a surface acoustic wave (SAW)
on the crystal surface. A SAW-induced AC electric field penetrates the sample
and interacts with charge carriers (holes).  As a result, the SAW attenuation
$\Gamma$ and velocity $v$ both acquire additional contributions, which can be
related to the complex conductance $\sigma$ with the following
formulae\cite{KaganSAWen}:
\begin{equation}
  \label{eq:SAWeqns}
  \begin{aligned}
    \Gamma &= 8.68q\frac{K^2}{2}A \frac{\Sigma_1}{\Sigma_1^2 + [\Sigma_2 + 1]^2}, \frac{\text{dB}}{\text{cm}}; \\
    \frac{\Delta v}{v_0} &= \frac{K^2}{2}A \frac{[\Sigma_2 + 1]}{\Sigma_1^2 + [\Sigma_2 + 1]^2}, \\
    \Sigma_i &= \frac{4\pi t(q)}{\varepsilon_S v_0}\sigma_i, i=1,2\\
    A &= 8b(q)[\varepsilon_1 + \varepsilon_0] \varepsilon_0^2 \varepsilon_S
    \exp\left(-2q(a + d)\right),
  \end{aligned}
\end{equation}
where $K^2$ is the electromechanical coupling constant for \mbox{LiNbO$_3$}; $q$
and $v_0$ are the SAW wave vector and velocity in \mbox{LiNbO$_3$},
respectively; $a$ is the vacuum gap between the \mbox{LiNbO$_3$} crystal and the
sample; $d$ is the distance between the sample surface and the QW layer;
$\varepsilon_1$, $\varepsilon_0$ and $\varepsilon_s$ are the dielectric
constants of \mbox{LiNbO$_3$}, of vacuum, and of the semiconductor,
respectively; $b$ and $t$ -- are complex functions of $q$, $a$, $d$,
$\varepsilon_0$, $\varepsilon_1$, and $\varepsilon_s$.

\begin{figure}[bt]
  \centering
  \includegraphics[width=8.5cm, keepaspectratio]{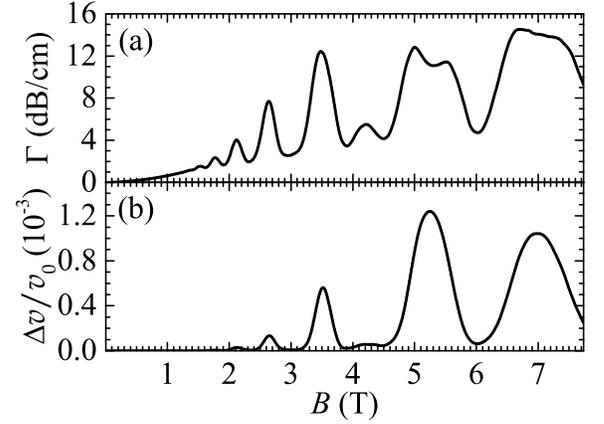}
  \caption{\label{fig:GammaVelocity} Dependence of the surface acoustic wave
  attenuation~(a) and velocity~(b) for the sample \mbox{6777-2} on the magnetic field;
  $f~=~30~\text{MHz}$, $T~=~1.8~\text{K}$.}
\end{figure}

\begin{figure}[bt]
  \centering
  \includegraphics[width=8.5cm, keepaspectratio]{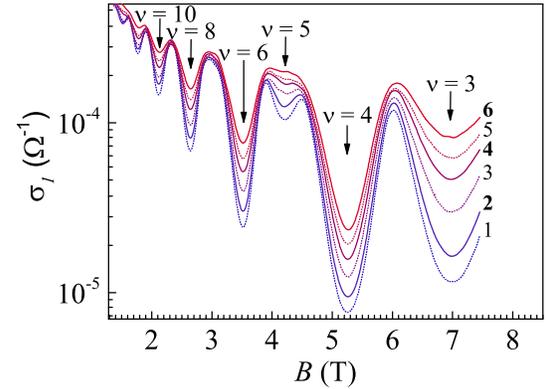}
  \caption{\label{fig:Sigma} Dependence of the real part of AC conductance of
  sample 6777-2 on the magnetic field at different temperatures:
  1~--~$1.8~\text{K}$,
  2~--~$2.1~\text{K}$,
  3~--~$2.7~\text{K}$,
  4~--~$3.2~\text{K}$,
  5~--~$3.7~\text{K}$,
  and
  6~--~$4.2~\text{K}$;
  $f = 30~\text{MHz}$. $\nu$ is the filling factor.}
\end{figure}

Measurements of the SAW attenuation $\Gamma$ and velocity $v$ have been
performed in magnetic fields up to $8~\text{T}$ perpendicular to the QW plane
and at temperatures from 1.7~to $4.2~\text{K}$. Fig.~\ref{fig:GammaVelocity}
shows $\Gamma$ and $v$ plotted as functions of magnetic field. Both curves
demonstrate oscillations. The real part $\sigma_1$ of the complex conductance
was then obtained from these measurements by solving
equations~\eqref{eq:SAWeqns}.  Fig.~\ref{fig:Sigma} shows the conductance
$\sigma_1$ of sample 6777-2 calculated in this way, as a function of magnetic
field at different temperatures.  The curves demonstrate Shubnikov -- de Haas
oscillations, which evolve into the characteristic oscillations of the integer
quantum Hall effect regime in stronger fields.  The same picture has been
observed with all samples.

\subsection{Determination of g-factor}
The analysis of the temperature dependence of the conductance at the minima of
the oscillations allows us to determine the $\mathrm{g}_{\perp}$-factor using
the following procedure. We have found that at each oscillation minimum there
is a temperature range where the conductance is characterized by a constant
activation energy $\Delta$:
\begin{equation}
  \sigma_1 \sim \exp\left(-\frac{\Delta}{2kT}\right).
\end{equation}
We have then determined $\Delta$ for each minimum by performing linear fitting
of the dependence of $\ln\sigma_1$ on $T^{-1}$.

\begin{figure}[thb]
  \centering
  \includegraphics[width=8.5cm, height=6cm, keepaspectratio]{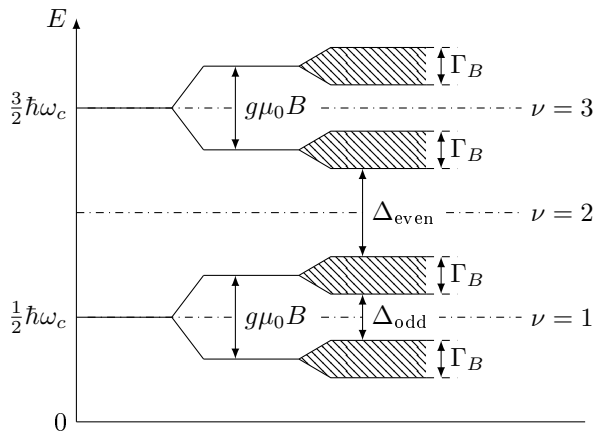}
  \caption{\label{fig:LandauLevels} Landau level structure at an oscillation
  minimum. $\nu$ is the filling factor. Hatched areas denote the Landau level
  broadening $\Gamma_B$.}
\end{figure}

On the other hand, one can write down expressions for activation energy
considering the Landau level structure at the minima of oscillations, where the
Fermi level is halfway between adjacent Landau levels. Even and odd filling
factors correspond to orbital and spin splitting of Landau levels, respectively,
as is shown in Fig.~\ref{fig:LandauLevels}. Taking the Landau level broadening
$\Gamma_B$ into account as $\Gamma_B = C\sqrt{B}$ (Ref.~\onlinecite{Coleridge}), one can express
activation energies for even and odd filling factors as
\begin{align}
  \Delta_{\text{odd}} &= \mathrm{g}_{\perp}\mu_0B - \Gamma_B,
  \label{eq:DeltaOdd} \\
  \Delta_{\text{even}} &= \hbar\omega_c - \mathrm{g}_{\perp}\mu_0B - \Gamma_B,
  \label{eq:DeltaEven}
\end{align}
where $\hbar\omega_c$ is the cyclotron energy, and $\mu_0$ is the Bohr magneton.
Equations~\eqref{eq:DeltaOdd} and~\eqref{eq:DeltaEven} are then solved for
$\mathrm{g}_{\perp}$ and $C$ for each $(\Delta_{\text{even}},
\Delta_{\text{odd}})$ pair.

\begin{table}[bt]
  \centering
  \caption{\label{tab:GFactorPairs} Values of the effective
$\mathrm{g}_{\perp}$-factor and the Landau level broadening factor $C$,
calculated for different pairs of conductance oscillation minima (sample
\mbox{6777-2}).}
  \begin{tabular}{| c c | c c |}
    \hline
    $\nu_{\text{odd}}$ &
    $\nu_{\text{even}}$ &
    \mbox{\hspace{1em}$\mathrm{g}_{\perp}$\hspace{1em}} &
    $C$ ($\text{meV}/\text{T}^{1/2}$) \\
    \hline
    3 & 6 & 8.5 & 0.8 \\
    3 & 8 & 8.5 & 0.8 \\
    3 & 10 & 8.3 & 0.8 \\
    3 & 12 & 8.4 & 0.8 \\
    5 & 6 & 8.4 & 0.8 \\
    5 & 8 & 8.4 & 0.8 \\
    5 & 10 & 8.2 & 0.8 \\
    5 & 12 & 8.2 & 0.8 \\
    \hline
  \end{tabular}
\end{table}

\begin{table}[bt]
  \centering
  \caption{\label{tab:GFactorResults} Values of the effective
  $\mathrm{g}_{\perp}$-factor obtained for different samples.}
  \begin{tabular}{| c c c |}
    \hline
    Sample & Hole density ($\text{cm}^{-2}$) & $\mathrm{g}_{\perp}$-factor \\
    \hline
    7989-2 & $3.9 \times 10^{11}$ & $10  \pm 2  $ \\
    6745-3 & $4.5 \times 10^{11}$ & $9.0 \pm 1.4$ \\
    8005   & $4.5 \times 10^{11}$ & $8.9 \pm 1.5$ \\
    6745-2 & $4.9 \times 10^{11}$ & $8.9 \pm 1.4$ \\
    6777-2 & $5.2 \times 10^{11}$ & $8.4 \pm 1.3$ \\
    K6016  & $6.0 \times 10^{11}$ & $6.9 \pm 1.1$ \\
    7974-2 & $6.2 \times 10^{11}$ & $7.3 \pm 1.2$ \\
    \hline
  \end{tabular}
\end{table}

Table~\ref{tab:GFactorPairs} shows the values obtained for sample
\mbox{6777-2}.  On can see from the table that the $\mathrm{g}_{\perp}$-factor
does not depend on the magnetic field within our experimental error (estimated
as $10$--$15\%$).

\begin{figure}[hbt]
  \centering
  \includegraphics[width=8.5cm, keepaspectratio]{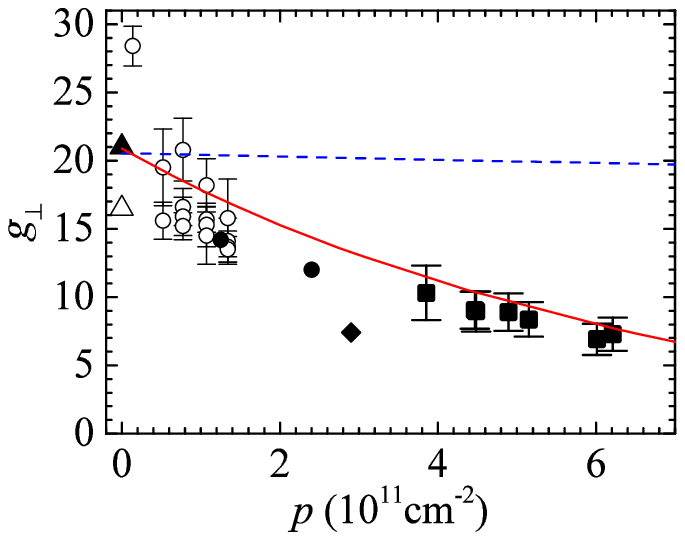}
  \caption{\label{fig:GFactorPlot} Density dependence of the $\mathrm{g}_{\perp}$-factor:
    ($\blacktriangle$)~--~g-factor of bulk Ge;
    ($\blacklozenge$)~--~Ref.~\onlinecite{ZudovFQHE}, the Landau level broadening is neglected;
    ({\large $\bullet$})~--~Refs.~\onlinecite{Arapov} and~\onlinecite{ContributionsGFactor};
    ($\vartriangle$)~--~Ref.~\onlinecite{Nenashev};
    ($\bigcirc$)~--~Ref.~\onlinecite{SmallSheetDensityGFactor};
    ($\blacksquare$)~--~the~present work.
    The dashed line shows the result of theoretical calculations
    based on the
$4\times 4$ $\mathbf{k}\cdot\mathbf{p}$ model\cite{NenashevPrivCorr}.
    The solid line shows the $\mathrm{g}_{\perp}$-factor evaluated with
formula~\eqref{eq:DurnevFormula} from the theoretical computation of the
effective mass in the $6\times 6$ $\mathbf{k}\cdot\mathbf{p}$ model.~\cite{HansRossner2007}
}
\end{figure}

Using the procedure described above, we have obtained the
values of $\mathrm{g}_{\perp}$-factor for every sample; the results are listed
in Table~\ref{tab:GFactorResults}.  It is worth mentioning that the effective
mass dependence on the hole density has been neglected in our calculations, and
a constant value of $m^{*} = 0.1m_0$, where $m_0$ is the free electron mass,
has been taken instead. This approximation is valid because the effective mass
changes only by $15\%$ within the range of carrier densities studied in the
present work.\cite{HansRossner2007}

\subsection{Discussion}
Fig.~\ref{fig:GFactorPlot} summarizes the results of several studies of the
$\mathrm{g}_{\perp}$-factor value in \mbox{SiGe/Ge/SiGe} structures, including
those of the present work.
We attribute the decrease in the $\mathrm{g}_{\perp}$-factor to valence band
non-parabolicity, which is due to the mixing of HH and LH states away from the
zone center. At the center of the valence band the $\mathrm{g}_{\perp}$-factor
of 2D holes should be approximately equal to the g-factor of heavy holes in
bulk Ge ($\mathrm{g} = 20.4$). With growing hole density, the
$\mathrm{g}_{\perp}$-factor is expected to decrease from the band-center value.

The reason is that in the presence of non-zero momentum component
$p_{\parallel}$ along the QW plane, the hole state is no longer the pure heavy
hole state $\left| \pm\frac{3}{2}\right\rangle$ but contains an admixture of
the light hole state $\left|\pm\frac{1}{2}\right\rangle$. The degree of this
admixture is roughly proportional to $p_{\parallel}^2$, which, at the Fermi
level, is proportional to the hole density. Thus, the
$\mathrm{g}_{\perp}$-factor becomes a combination of the
$\mathrm{g}_{\perp}$-factors of heavy and light holes, taken with weights which
are proportional to fractions of heavy and light holes in the observed
state.\cite{NenashevPrivCorr}
The result of a calculation\cite{NenashevPrivCorr} based on the $4 \times 4$
$\mathbf{k}\cdot\mathbf{p}$ Hamiltonian in the infinite well approximation is shown with the blue
dashed line in Fig.~\ref{fig:GFactorPlot}. As is seen from the figure, this
theoretical curve fails to describe the experimental results.

Several numerical calculations that were based on a more rigorous model
with the $6 \times 6$ $\mathbf{k}\cdot\mathbf{p}$ Hamiltonian have been made to explain the
density dependence of the effective mass.\cite{HansRossner2007,StrainedSiGeCRTheory}  In these works,
along with the split-off (SO) band, self-consistent potential profiles and the effects of
strain due to the lattice mismatch between Ge and SiGe were also taken into
account. The results of these computations show good agreement with the
experimental dependence of the effective mass on density. The structures
studied in Ref.~\onlinecite{HansRossner2007} have almost the same parameters as
those investigated in the present work, so we used the results of calculations
from Ref.~\onlinecite{HansRossner2007} to interpret the
$\mathrm{g}_{\perp}$-factor dependence on density as follows.

As the density dependencies of both effective mass and
$\mathrm{g}_{\perp}$-factor of heavy holes have the same microscopic nature,
they can be related to each other as follows\cite{Wimbauer1994}:
\begin{equation}
\label{eq:DurnevFormula}
|\mathrm{g}_{\perp}| = \left|2\left(
  -3\mathcal{K} + (\gamma_1 + \gamma_2) - \frac{m_0}{m^{*}}
\right)\right|,
\end{equation}
where $m_0$ is the free electron mass, $\gamma_1 = 13.38$, $\gamma_2 = 4.26$
and $\mathcal{K} = 3.41$ are the Luttinger parameters of
Ge.~\cite{IvchenkoPikus} The dependence of the $\mathrm{g}_{\perp}$-factor on
density was evaluated from the theoretical dependence of the effective mass on
density\cite{HansRossner2007} by means of formula~\eqref{eq:DurnevFormula}. The
result is shown with the solid line in Fig.~\ref{fig:GFactorPlot}. As is
seen, it describes the experimental data with good agreement. Thus, the
observed experimental $\mathrm{g}_{\perp}$-factor dependence on density can
be attributed to the interaction between HH, LH and SO subbands, as
affected by the strain present in the quantum well.

\section{Conclusions}
We have performed contactless conductance measurements on the 2D hole gas in
\mbox{p-SiGe/Ge/SiGe} structures by means of acoustic spectroscopy, and
determined the effective $\mathrm{g}_{\perp}$-factor. It was found that the
$\mathrm{g}_{\perp}$-factor depends approximately linearly on the hole density,
but is independent of the magnetic field, the Si content $x$ in the spacer
layers, and the character of the modulation doping (see
Table~\ref{tab:Samples}). The observed change is attributed to valence band
non-parabolicity.

\begin{acknowledgements}
  We are grateful to A.~V.~Nenashev, S.~A.~Tarasenko, M.~V.~Durnev, and
  N.~G.~Shelushinina for useful discussions of the experimental results.  Author
  A.~A.~D. also acknowledges the support by the~Government of the~Russian
  Federation (Grant 074-U01).
\end{acknowledgements}


%

\end{document}